\documentstyle[11pt,moriond,epsfig]{article}
\newcommand{\nc}{\newcommand}
\nc{\hc}{\hbox {h.c.}}
\nc{\re}{\hbox {Re}}
\nc{\im}{\hbox {Im}}
\nc{\etal}{\hbox{et al.}}
\nc{\ra} {\rightarrow}
\nc{\cw}{\cos\theta_W}        \nc{\sw}{\sin\theta_W}
\nc{\ttbar}{t\bar{t}}
\nc{\bbbar}{b\bar{b}}
\nc{\tanb} {\tan \beta}
\nc{\twbdec} {t\rightarrow W^+ b}
\nc{\tbwbdec} {\bar{t} \rightarrow W^- \bar{b}}
\nc{\hprod} {e^+e^- \ra Z^\ast \ra H Z}
\nc{\epem} {e^+e^-}
\nc{\wpwm} {W^+W^-}
\nc{\tbar} {\bar{t}}
\nc{\bbar} {\bar{b}}
\nc{\wpp} {W^+}
\nc{\mt}{m_t}
\nc{\mts}{m_t^2}
\nc{\mw} {m_W}
\nc{\mws} {m_W^2}
\nc{\mz} {m_Z}
\nc{\mzs} {m_Z^2}
\nc{\mh} {m_H}
\nc{\mhs} {m_H^2}
\nc{\ma} {m_A}
\nc{\mas} {m_A^2}
\nc{\hdec}{H \ra t\bar{t}}
\nc{\ttbardec}{\ttbar \ra W^+W^-\bbbar}
\nc{\po}{\Phi_1}
\nc{\pod}{\Phi_1^\dagger}
\nc{\pht}{\Phi_2}
\nc{\phtd}{\Phi_2^\dagger}
\nc{\phtt}{{\tilde{\Phi}}_2}
\nc{\popo}{\po^\dagger\po}
\nc{\phtpt}{\pht^\dagger\pht}
\nc{\popt}{\po^\dagger\pht}
\nc{\phtpo}{\pht^\dagger\po}
\nc{\sq}{\sqrt{2}}
\nc{\nsd} {N_{SD}}
\nc{\ntt} {N_{tt}}
\nc{\vs}{\vspace{2mm}}
\nc{\sty}{\hat{S}^t_1} \nc{\pty}{\hat{P}^t_1}
\nc{\sts}{(\sty)^2}      \nc{\pts}{(\pty)^2}
\nc{\yts}{\sts+\pts}
\nc{\sby}{\hat{S}^b_1} \nc{\pby}{\hat{P}^b_1}
\nc{\sbs}{(\sby)^2}      \nc{\pbs}{(\pby)^2}
\nc{\ybs}{\sbs+\pbs}

\def\ie{{\it i.e.}}

\def\sb{s_\beta}
\def\cb{c_\beta}
\def\rts{\sqrt s}

\def\hsm{h_{\rm SM}}
\def\mhsm{m_{\hsm}}

\def\h{h}

\def\lsim{\mathrel{\raise.3ex\hbox{$<$\kern-.75em\lower1ex\hbox{$\sim$}}}}
\def\gsim{\mathrel{\raise.3ex\hbox{$>$\kern-.75em\lower1ex\hbox{$\sim$}}}}

\def\Journal#1#2#3#4{{#1} {\bf #2}, #3 (#4)}

\def\PLB{{\em Phys. Lett.}  B}
\def\PRL{\em Phys. Rev. Lett.}
\def\PRD{{\em Phys. Rev.} D}
\def\ZPC{{\em Z. Phys.} C}

\def\ra{\rightarrow}

\def\be{\begin{equation}}
\def\ee{\end{equation}}
\def\bea{\begin{eqnarray}}
\def\eea{\end{eqnarray}}

\begin{document}
\hfill IFT/99-11

\hfill hep-ph/9906336

\vspace*{4cm}
\title{FINDING CP-VIOLATING HIGGSES}

\author{ J. KALINOWSKI }

\address{Instytut Fizyki Teoretycznej, Uniwersytet Warszawski\\
Ho\.za 69, 00681 Warsaw, Poland}

\maketitle\abstracts{In a general two-Higgs-doublet model with CP
violation in the Higgs sector, the three neutral physical Higgs 
bosons have no definite CP 
properties.  A new sum rule relating Yukawa
and Higgs--$Z$ couplings  implies that a neutral Higgs
boson cannot escape detection at an $e^+e^-$ collider 
if it is kinematically accessible in $Z$+Higgs, $b\bar b+$Higgs
and $t\bar t+$Higgs production, irrespective of the mixing
angles and the masses of the other neutral Higgs bosons.
The implications of the sum rules
for Higgs discovery at the Tevatron and LHC are briefly mentioned.}

\newpage
\section{Introduction}
The origins of the electroweak symmetry breaking and 
the CP violation are still not
understood. In the Standard Model (SM) the former is achieved
spontaneously by the non-zero vacuum expectation value of the Higgs
field, while the latter is parametrized by the complex Yukawa
Higgs-fermion couplings.  Although such a minimal model of the
electroweak interactions describes exceedingly well a wealth of
experimental data, the SM cannot be considered as a fundamental theory
since neither the structure of the model, nor its parameters are
predicted. They are merely built in.

The models of mass generation by elementary scalars predict one (like
in the SM) or more physical Higgs bosons. Already the simplest
extension of the SM with two scalar Higgs doublets (2HDM)
predicts the existence of 5 physical Higgs bosons: three electrically
neutral and a charged pair. The CP-conserving (CPC) version of the
2HDM has received a considerable attention, especially in the context
of the minimal supersymmetric model (MSSM). It predicts two neutral
Higgs bosons ($h^0$ and $H^0$) to be CP-even and one CP-odd neutral
state ($A^0$). However, such a distinction may get lost beyond the
Born approximation if the soft SUSY breaking parameters have a nonzero
CP-violating phases~\cite{cpp}. 
As a result, all three neutral Higgs states may
mix and the mass eigenstates $h_1$, $h_2$ and $h_3$ will have no definite
CP properties.

In a general (non-supersymmetric) 2HDM the Higgs sector itself may also
generate CP violation (CPV). The possibility that CP violation,
spontaneous and/or explicit, derives largely (or entirely) from the
Higgs sector is particularly appealing~\cite{weinberg}.  In the
CPV 2HDM the neutral Higgs bosons mix already at the tree level. From
the phenomenological point of view, a critical question then arises whether
the additional freedom in Higgs boson couplings is sufficient to
jeopardize our ability to find light neutral Higgs bosons.
We will see that the unitarity of the model 
implies a number of interesting sum rules for the Higgs-gauge boson 
\cite{mp,gghk}
and Higgs-fermion \cite{ggk}    couplings that 
guarantee the discovery in $e^+e^-$ collisions of any neutral Higgs boson
that is sufficiently light to be kinematically accessible in 
(a) the Higgs-strahlung, (b) the Higgs-pair production or (c) the 
 $t\bar t$+Higgs processes. 

\section{Sum rules for the Higgs boson couplings}

We consider the CPV 2HDM of electroweak interactions 
with two SU(2) Higgs doublets  
$\Phi_1=(\phi_1^+,\phi_1^0)$ and $\Phi_2=(\phi_2^+,\phi_2^0)$ 
defined \cite{ggk} by the potential 
\begin{eqnarray}
V(\po,\pht)&=&-\mu_1^2\popo-\mu_2^2\phtpt  
+\lambda_1(\popo)^2+\lambda_2(\phtpt)^2+
\lambda_3(\popo)(\phtpt)\nonumber \\
&&+\lambda_4|\popt|^2+
[\lambda_5(\popt)^2+\hc] -[\mu_{12}^2\popt+\hc]
\end{eqnarray}
and the Yukawa couplings
\begin{equation}
{\cal L}_Y=-(\bar{u}_i,\bar{d}_i)_L \Gamma_u^{ij} \phtt {u_j}_R 
           -(\bar{u}_i,\bar{d}_i)_L \Gamma_d^{ij} \po  {d_j}_R
           -(\bar{\nu}_i,\bar{e}_i)_L \Gamma_e^{ij} \po  {e_j}_R + \hc,
\label{yukcoupl}
\end{equation}
where $i,j$ are generation indices and $\phtt$ is defined as 
$i\sigma_2 \pht^\ast$.

If $\im({\mu_{12}^\ast}^4\lambda_5)\neq 0$, the CP is violated explicitly. 
When $\im({\mu_{12}^\ast}^4\lambda_5)= 0$,  but
$|\mu_{12}^2/ 2\lambda_5v_1v_2|<1$, CP is spontaneously violated 
since the minimum of the potential occurs  for 
$<\po>=v_1/\sqrt{2}$  (without loss of generality, $v_1$ 
is positive) and   
$<\pht>=v_2e^{i\theta}/\sqrt{2}$, where $\cos\theta=\mu_{12}^2/ 
2\lambda_5v_1v_2$.
In this normalization $v\equiv 
\sqrt{v_1^2+v_2^2}=2m_W/g=246\,\mbox{GeV}$.

After SU(2)$\times$U(1) gauge symmetry breaking, the state 
$\sqrt2(\cb\mbox{Im}\phi_1^0+ \sb\mbox{Im}\phi_2^0)$ 
becomes a would-be Goldstone boson which is absorbed in giving 
mass to the $Z$ gauge boson.
(Here, we use the notation $\sb\equiv\sin\beta$, $\cb\equiv\cos\beta$,
where $\tanb=v_2/v_1$.)
The remaining three neutral degrees of freedom $
(\varphi_1,\varphi_2,\varphi_3)\equiv
\sqrt 2(\mbox{Re}\phi_1^0, \, \mbox{Re}\phi_2^0, \,
 s_\beta\mbox{Im}\phi_1^0-c_\beta\mbox{Im}\phi_2^0) $
are not mass eigenstates. The physical neutral Higgs bosons $h_i$
($i=1,2,3$) are obtained by an orthogonal transformation, $h=R
\varphi$, where the rotation matrix is given in terms of three Euler
angles ($\alpha_1, \alpha_2,\alpha_3$) by
\begin{eqnarray} 
R=\left(\begin{array}{ccc}
  c_1     &  -s_1c_2          &     s_1s_2  \\
  s_1c_3  & c_1c_2c_3-s_2s_3  &  -c_1s_2c_3-c_2s_3\\
  s_1s_3 & c_1c_2s_3+s_2c_3 & -c_1s_2s_3+c_2c_3 \end{array}\right),
\label{mixing}
\end{eqnarray}
where $s_i\equiv\sin\alpha_i$ and $c_i\equiv\cos\alpha_i$.
Without loss of generality, we assume $m_{h_1}\le m_{h_2} \le m_{h_3}$. 
The Yukawa interactions of the $h_i$ mass-eigenstates are not
invariant under CP
\begin{equation} 
{\cal L}=h_i\bar{f}(S^f_i+iP^f_i\gamma_5)f, \label{coupl} 
\end{equation}
where the scalar ($S^f_i$) and pseudoscalar ($P^f_i$) couplings are
functions of the mixing angles. For up-type ($f=u$) and down-type ($f=d$) 
quarks we have 
\begin{equation}
S^u_i=-\frac{m_u}{v s_\beta}R_{i2},\;\;
P^u_i=-\frac{m_u}{v s_\beta}c_\beta R_{i3}, \;\; \mbox{\rm and}\;\; 
S^d_i=-\frac{m_d}{v c_\beta}R_{i1},\;\;
P^d_i=-\frac{m_d}{v c_\beta}s_\beta R_{i3}\,,
\label{absd}
\end{equation}
and similarly for charged leptons. 

The absence of any $\epem\to Z \hsm$ signal in LEP data  translates into
a lower limit on $\mhsm$. The latest analysis of four LEP experiments
at $\sqrt{s}$ up to 189 GeV 
implies $\mhsm$ greater than about 90 GeV \cite{felcini}. 
The negative results of Higgs
boson searches at LEP can be formulated as restrictions on the
parameter space of the 2HDM and more general Higgs sector models. As has
been shown in Refs.~\cite{mp,gghk}, the sum rules for the Higgs--$Z$ boson
couplings derived in the CP-conserving 2HDM can be generalized to the
CP-violating case 
to yield a sum rule 
\begin{equation}
  \label{oldsr}
C_{i}^2+C_{j}^2+C_{ij}^2=1 , 
\end{equation}
where $i\neq j$ are any two of the three possible
indices, and $C_{i}$ and 
$C_{ij}$ denote the reduced $ZZh_i$ and $Zh_ih_j$ couplings
\begin{eqnarray}
C_i& \equiv & (\frac{g m_Z}{c_W})^{-1} g_{ZZh_i} = 
s_{\beta} R_{i2}+c_{\beta}R_{i1},
\label{zzhcoup}  \\
C_{ij}& \equiv & (\frac{g}{2c_W})^{-1} g_{Zh_ih_j} =
(s_{\beta}R_{i1}-c_{\beta}R_{i2}) R_{j3} - 
(s_{\beta}R_{ij}-c_{\beta}R_{j2}) R_{i3}.
\label{zhhcoup}
\end{eqnarray}
The power of Eq.~(\ref{oldsr})
with $i,j=1,2$ for LEP physics derives from two facts: it
involves only two of the neutral Higgs bosons; and the
experimental upper limit on any one $C_i^2$ derived from $\epem\to
Z\h_i$ data is very strong: $C_i^2\lsim 0.1$ for $m_{\h_i}\lsim 70$ GeV.
Thus, if $\h_1$ and $\h_2$ are both below about $70$ GeV in mass, then
Eq.~(\ref{oldsr}) requires that $C_{12}^2\sim 1$, whereas for such
masses the limits on $\epem\to \h_1\h_2$ from LEP2 data require
$C_{12}^2\ll 1$. As a result, there cannot be two light Higgs bosons
even in the general CP-violating case; the excluded region in the
$(m_{\h_1},m_{\h_2})$ plane that results from a recent analysis by the
DELPHI Collaboration is quite significant~\cite{cpvdelphi}.

\begin{figure}[ht]
\begin{center}
\psfig{figure=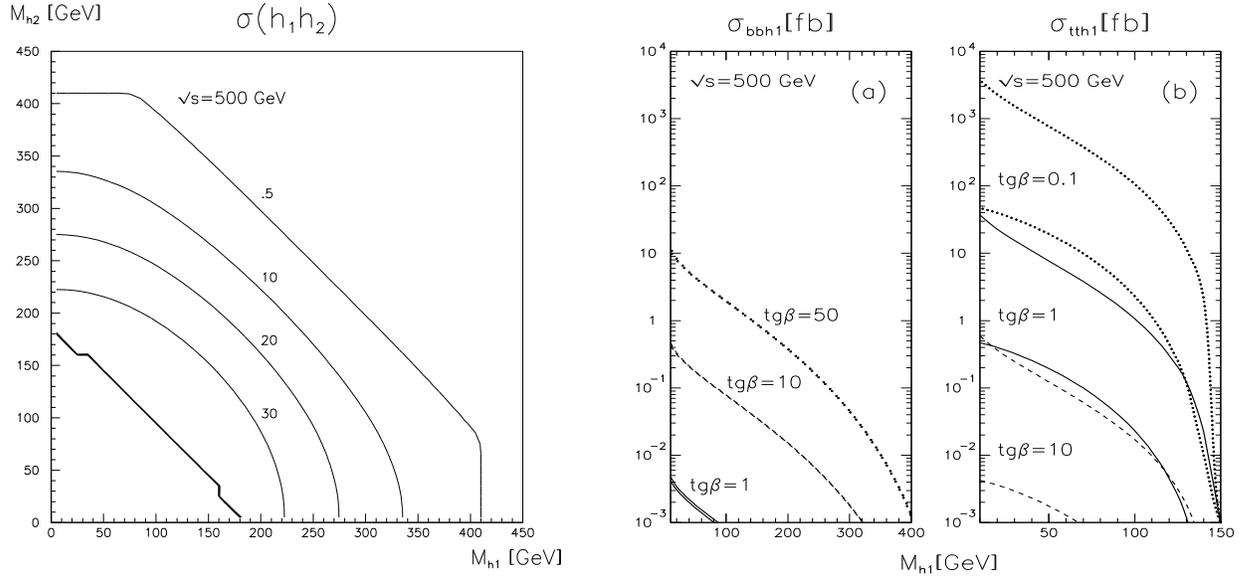,height=3in}  
\end{center}
\caption{{\it Left panel:} Contour lines for 
${\rm min}[\sigma(e^+e^-\rightarrow h_1 h_2)]$ in units of fb's. 
The
contour lines are plotted for $\tanb=0.5$; the plots are
virtually unchanged for larger values of $\tanb$. The contour lines 
overlap in the inner corner as a result of
excluding mass choices inconsistent with experimental 
constraints from LEP2 data.
{\it Right panel:}  The minimal and maximal values 
of the cross sections for $e^+e^- \rightarrow 
\bbbar h_1$ (a) and $e^+e^-\rightarrow \ttbar h_1$ (b),
the same type of line (dots for $\tanb=0.1$ and $t\bar t h_1$, solid
for $\tanb=1$, dashes for $\tanb=10$, dots for $\tanb=50$ and $b\bar bh_1$)
is used for the minimal and maximal values of the cross sections.
In the case of $\bbbar h_1$, the
minimal and maximal values of the cross sections are almost the same.
Masses of the remaining Higgs bosons are assumed to be $1000$ GeV.
\label{fig1}}
\end{figure}

If even one of the three processes, $Zh_1$, $Zh_2$ (Higgs-strahlung)
and $h_1h_2$ (pair production), is beyond the collider's kinematical
reach, the sum rule in Eq.~(\ref{oldsr}) is not sufficient to
guarantee $h_1$ or $h_2$ discovery even if there is a light $h_i$.
For example, with $C_{12}\sim 1$ and $C_{1,2} \sim 0$ 
satisfying eq.~(\ref{oldsr}), the 
$h_1h_2$ production could be kinematically closed while  
$Zh_1$ and $Zh_2$ production would be suppressed
by $C_1$ and $C_2$.  
However, in this case we can exploit another sum rule derived in \cite{ggk}
which constrain the Yukawa and $ZZ$ couplings of any one Higgs boson, namely 
(for obvious reasons we consider the third generation fermions)  
\begin{eqnarray}
  \label{yuksr}
(\hat{S}^t_i)^2 + (\hat{P}^t_i)^2 
&=&\left(\frac{\cos\beta}{\sin\beta}\right)^2 \left[
1+\frac{C_i}{\cos^2\beta}(2 \hat{S}^b_i \cos^2\beta+ C_i)
\right]\,;
\nonumber \\
(\hat{S}^b_i)^2 + (\hat{P}^b_i)^2 
&=&\left(\frac{\sin\beta}{\cos\beta}\right)^2 \left[
1+\frac{C_i}{\sin^2\beta}(2 \hat{S}^t_i \sin^2\beta+ C_i)
\right]\,.
\end{eqnarray}
where $\hat{S}^f_i\equiv S^f_i v/m_f$, $\hat{P}^f_i\equiv 
P^f_i v/m_f$. Combining the two sum rules we get 
\begin{eqnarray}
  \label{finalsr}
  \sin^2\beta [(\hat{S}^t_i)^2 + (\hat{P}^t_i)^2]
+ \cos^2\beta [(\hat{S}^b_i)^2 + (\hat{P}^b_i)^2]=1  
\end{eqnarray}
independently of $C_i$. Eq.~(\ref{finalsr})
implies that the Yukawa couplings to top and bottom
quarks cannot be simultaneously suppressed, {\it i.e.} if $C_i\sim 0$
at least one $h_i$ Yukawa coupling must be large.
The complete
Higgs hunting strategy at $e^+e^-$ colliders, and at hadron colliders
as well, should therefore include not only the Higgs-strahlung 
and Higgs-pair production but also the Yukawa processes~\footnote{The
importance of the Yukawa processes in the context of a CP conserving
2HDM for large $\tan\beta$ has been stressed in the past many times
\cite{oldyukawa,dkz}.} with Higgs radiation off top and
bottom quarks in the final state.

\section{ Higgs boson production in $e^+e^-$ colliders}

In order to treat the three $h_1$ production mechanisms: 
(a) $e^+e^-\ra Zh_1$, (b)  $e^+e^-\ra h_1h_2$, and (c) the 
Yukawa processes $e^+e^-\ra f\bar{f}h_1$,  on the same footing, 
we consider the   $ f\bar{f}h_1$ final state at future $e^+e^-$ colliders. 
The  processes (a) and (b) contribute to this final state 
when $Z\to f\bar f$ and $h_2\to f\bar f$, respectively.
Since all fermion and Higgs boson masses in the final state must be
kept nonzero, the formulae for the cross section are quite involved.
They can be found in \cite{ggk}. 

If the coupling of the $h_1$ to the $Z$ boson is not dynamically
suppressed, \ie\ $C_1$ is substantial, then the Higgs-strahlung
process, $e^+e^-\rightarrow Zh_1$, will be sufficient to find it. In
the opposite case, which is a main focus of my talk, one has to
consider the other processes (b) and/or (c), for which the sum rules
(\ref{oldsr}) and (\ref{yuksr}) will imply that the neutral Higgs
boson(s), if kinematically accessible, will be produced at a
comfortably high rate at a high luminosity future linear $e^+e^-$
collider\cite{lc}. For definiteness we will take an $e^+e^-$ collider with
$\sqrt{s}=500$ GeV and integrated luminosity $L=500$ fb$^{-1}$
and assume that $C_i$ are suppressed so that fewer than 50
$Zh_i$ events are expected.  We will consider two situations:
\begin{description}
\item{(i)} two light Higgs bosons:
{\it i.e.} $m_{h_1}+m_{h_2},m_{h_1}+\mz,m_{h_2}+\mz<\sqrt{s}$.  If
$C_1,\, C_2\ll1$, then from Eq.~(\ref{oldsr}) it follows that Higgs
pair production is at full strength, $C_{12}\sim 1$. In the left panel
of Fig.1 contour lines are shown for the minimum value of the pair
production cross section, $\sigma(e^+e^- \rightarrow h_1 h_2)$ as a
function of Higgs boson masses.  The mimimum of
$\sigma(h_1h_2)$ is found by scanning over the mixing angles
$\alpha_i$ consistent with present experimental constraints 
on $C_i$ (which
roughly exclude $m_{h_1}+m_{h_2}\lsim 180$ GeV) and the above
assumption of less that 50 $Zh_i$ events.  With $L=500$ fb$^{-1}$ a
large number of events (large enough to allow for selection cuts and
experimental efficiencies) is predicted for a broad range of Higgs
boson masses.  If 50 events before cuts and efficiencies prove
adequate (i.e. $\sigma>0.1$ fb), 
one can probe reasonably close to the kinematic boundary
defined by requiring that $m_{h_1}+\mz$, $m_{h_2}+\mz$ and
$m_{h_1}+m_{h_2}$ all be less than $\sqrt s$.

\item{(ii)} one light Higgs boson:
{\it i.e.} $m_{h_1}+\mz<\sqrt s$, $m_{h_1}+m_{h_2},m_{h_2}+\mz>\sqrt{s}$. 
If $C_1$
is small the sum rules (\ref{yuksr}) imply that Yukawa
couplings may still allow detection of the $h_1$. This is 
demonstrated in the right panel of Fig.~\ref{fig1}, where 
the minimum and maximum values of
$\sigma(e^+e^- \rightarrow f\bar{f}h_1)$ for $f=t$ (a) and $f=b$ (b)
are drawn as  functions of the Higgs boson mass.   It is seen 
that if $m_{h_1}$ is not large there will be sufficient events in 
either the $b\bar{b}h_1$ or the
$t\bar{t}h_1$ channel (and perhaps both) to allow $h_1$ discovery.
The smallest reach in $m_{h_1}$ arises if
$1\lsim\tanb\lsim 10$ and the $\alpha_i$'s are such that
the $t\bar t h_1$ cross section is minimal. Taking again 50 events 
as the observability criteria,   
at $\tanb=1$ we have $\sigma(b\bar b h_1)\ll 0.1$ fb for all $m_{h_1}$ 
while $\sigma_{\rm min}(t\bar t h_1)$ falls below $0.1$ fb for
$m_{h_1}>70$ GeV. At $\tanb=10$, $\sigma_{\rm min}(t\bar t h_1)\ll 0.1$ fb
and $\sigma_{\rm min}(b\bar b h_1)\simeq \sigma_{\rm max}(b\bar b h_1)$
falls below $0.1$ fb for $m_{h_1}>80$ GeV.
A $\rts=1$ TeV machine would considerably extend this mass reach.
\end{description}

The generalization to models with additional Higgs singlets
modifies the sum rules. Each singlet field introduces two more
physical neutral Higgs bosons which do not couple to $Z$ or to
fermions. As a result, in the sum rule (\ref{finalsr}), the factor 1 in
the RHS is replaced by the ``two-duoblet content'' of the $h_i$. At
least $1+2N_{\rm singlet}$ of the neutral Higgs bosons must be light
in order to guarantee\cite{ggk} that at least one of them will be observed in
$t\bar t h_i$ or $b\bar b h_i$ associated production.

\section{Conclusions}

The new sum rule, Eq.~(\ref{yuksr}),
relating the Yukawa and Higgs-$Z$ couplings of a general CP-violating
two-Higgs-doublet model implies 
that any one of the three neutral
Higgs bosons that is light enough to be produced in $\epem\to t\bar t h$
(implying that $\epem\to Zh$ and 
$\epem\to b\bar b h$ are also kinematically allowed) will
be found at an $e^+e^-$ linear collider of sufficient luminosity.
These same guarantees for the CPV 2HDM 
model do not apply to the Tevatron and LHC hadron colliders. 
In the case of the Tevatron, the small rate for $t\bar t$+Higgs
production is a clear problem \cite{tev2}. In the case of the LHC, 
a detailed study would be appropriate.  However,
existing studies in the context of supersymmetric
models can be used to point to parameter regions
that are problematical because of large backgrounds and/or signal dilution
due to sharing of available coupling strength.
Still, it is clear that the sum rules are sufficiently powerful
to imply that such parameter regions are of fairly limited extent.

\section*{Acknowledgments}
This work has been supported in part by the KBN Grant No. 
2~P03B~030~14, and by the Maria Sklodowska-Curie Joint Fund II
(Poland-USA) under grant No. MEN/NSF-96-252.

\end{document}